\documentclass[prl,twocolumn,showpacs,superscriptaddress]{revtex4-1}

\usepackage{amsmath}
\usepackage{amssymb}
\usepackage{graphicx}
\usepackage{longtable}%
\usepackage{dcolumn}
\usepackage{bm}
\usepackage{color}
\begin{document}
\title{
Emergence of an incipient ordering mode in FeSe\\} 
\author{Sahana R\"o{\ss}ler}
\email{roessler@cpfs.mpg.de}
\affiliation{Max Planck Institute for Chemical Physics of Solids,
N\"othnitzer Stra\ss e 40, 01187 Dresden, Germany}
\author{Cevriye Koz}
\affiliation{Max Planck Institute for Chemical Physics of Solids,
N\"othnitzer Stra\ss e 40, 01187 Dresden, Germany}
\author{Lin Jiao}
\affiliation{Max Planck Institute for Chemical Physics of Solids,
N\"othnitzer Stra\ss e 40, 01187 Dresden, Germany}
\author{Ulrich K. R\"o{\ss}ler}
\affiliation{IFW Dresden, Postfach 270016, 01171 Dresden, Germany}
\author{Frank~Steglich}
\affiliation{Max Planck Institute for Chemical Physics of Solids,
N\"othnitzer Stra\ss e 40, 01187 Dresden, Germany}
%
%
%
\author{Ulrich Schwarz}
\affiliation{Max Planck Institute for Chemical Physics of Solids,
N\"othnitzer Stra\ss e 40, 01187 Dresden, Germany}
\author{Steffen Wirth}
\email{wirth@cpfs.mpg.de}
\affiliation{Max Planck Institute for Chemical Physics of Solids,
N\"othnitzer Stra\ss e 40, 01187 Dresden, Germany}


\date{\today}

	\begin{abstract}
	
The structurally simplest Fe-based superconductor FeSe with a critical temperature $T_{c}\approx$ 8.5~K displays a breaking of the four-fold rotational symmetry at a temperature $T_{s}\approx 87~$K. 
We investigated the electronic properties of FeSe using scanning tunneling microscopy/spectroscopy (STM/S), magnetization, and electrical transport measurements. 
The results indicated two new energy scales (i) $T^{*} \approx$ 75~K denoted by an onset of electron-hole asymmetry in STS, enhanced spin fluctuations, and increased positive magnetoresistance; (ii) $T^{**} \approx$ 22 - 30~K, marked by opening up of a partial gap of about 8 meV in STS and a recovery of Kohler's rule. Our results reveal onset of an incipient ordering mode at $T^{*}$ and its nucleation below $T^{**}$. The ordering mode observed here, both in spin as well as charge channels, suggests a coupling between the spins with charge, orbital or pocket degrees of freedom.

\end{abstract}

\maketitle 

Soon after the discovery of superconductivity in LaFeAsO$_{1-x}$F$_{x}$ with a transition temperature $T_{c} \approx $ 26~K 
\cite{Kam2008}, several other superconducting iron-based materials have been found \cite{Joh2010,Ste2011,Hsu2008}. The maximum $T_{c}$ achieved so far in bulk samples of iron pnictides is  56 K \cite{Ren2008, Wan2008}, second to only the  high-$T_{c}$  cuprates. The generic phase diagrams of Fe-based superconductors (Fe-SC) and the  high-$T_{c}$  cuprates are remarkably similar; in both cases superconductivity emerges when the long-range antiferromagnetism is suppressed. While the pairing mechanism for high-$T_{c}$  superconductivity still remains elusive \cite{Nor2011}  in both families, spin-fluctuations are considered as one of the essential ingredients. The symmetry of the superconducting order parameter in the case of Fe-SC appears to be extended (possibly nodal) $s$-wave \cite{Maz2008,Han2010,Kas2014}, but other types of symmetries specific to the individual materials have also been considered \cite{Hir2011}.

Another feature ubiquitously found in the phase diagrams of cuprates, is the \textit{pseudogap}~(PG) state. In contrast, parent Fe-based superconductors display a structural phase transition at a temperature  $T_{s}$, which is usually accompanied by a spin density wave (SDW) order at temperature $T_{N} \leq T_{s}$. Also, the structural transition spontaneously  breaks the rotational symmetry of the underlying electronic system by lifting the degeneracy of the $d_{xz}/d_{yz}$ orbitals on Fe. In the cases where $T_{N}<T_{s}$, the resulting low-symmetry, orbitally ordered state in the temperature range  $T_{N}<T<T_{s}$ is referred to as a \textit{nematic state} by preferring Ising spin-fluctuations as a main driving force of the structural phase transition, albeit the origin of this nematic state is a matter of intense debate \cite{Fer2014}. 

 However, FeSe, the structurally simplest member of the Fe-SC (Fig. 1(a)), poses challenges to the spin-nematic picture. This is because it exhibits a structural transition from a tetragonal  $P \mathrm{4}/nmm$  to an orthorhombic $Cmma$ phase at $T_{s}\approx$~87~K \cite{Mc2009,Koz2014} and a superconducting transition at $T_{c}\approx$~8.5~K  \cite{Hsu2008}, but does not order magnetically down to $T=0$. Nonetheless, spin fluctuations have been detected in nuclear magnetic resonance (NMR) \cite{Baek2014,Bom2015,Imai2009} and neutron scattering experiments \cite{Rahn2015,Wang2015}. At $T_{s}$, a spontaneous orbital ordering \cite{Baek2014} and softening of the shear modulus has been observed \cite{Bom2015,Wat2015}. Several angle-resolved photo-emission spectroscopy (ARPES) measurements \cite{Mal2014,Shim2014,Naka2014,Wat2015,Zhang2015,Suzuki2015} report one small hole pocket at the center ($\Gamma$-point) of the Brillouin zone and one electron pocket at the $M$-point, but the interpretations of the temperature evolution of the band structure determined by ARPES do not agree with each other. Quantum oscillation measurements \cite{Wat2015,Tera2014,Wat2015a} are in agreement with the ARPES results concerning the hole pocket, but the electron pocket detected in these experiments seems not the same as the one observed in ARPES \cite{Wat2015}. Based on their results Watson $et~al.$ \cite{Wat2015,Wat2015a} suggested a scenario with two electron pockets, but only one hole pocket at the Fermi level $E_\mathrm{F}$. Similar conclusions were drawn from a mobility spectrum analysis of transport data, in which, in addition to compensated hole and electron type carriers, additional ultra-fast electrons with Dirac cone-like dispersion were detected \cite{Huy2014}. These results suggest that several competing interactions result in a complex Fermi surface below $T_{s}$ in FeSe.  

Here we investigate FeSe single crystals using scanning tunneling microscopy/spectroscopy (STM/S), magnetization, and electrical transport measurements. The results show evidence for two new energy scales  $T^{*}$ and  $T^{**}$  below $T_{s}$.
Our results indicate that $T^{*}$ represents the onset of an incipient order associated with enhanced spin fluctuations. Static nucleation of this mode below a second temperature $T^{**}$ appears to lead to a coupling between the spins with electronic charge, orbital, or pocket degrees of freedom as this temperature is discernible in anomalies of transport data and STM spectra. 

\begin{figure}[t]


\begin{minipage}{0.32\textwidth}
\includegraphics[width=3 cm,clip]{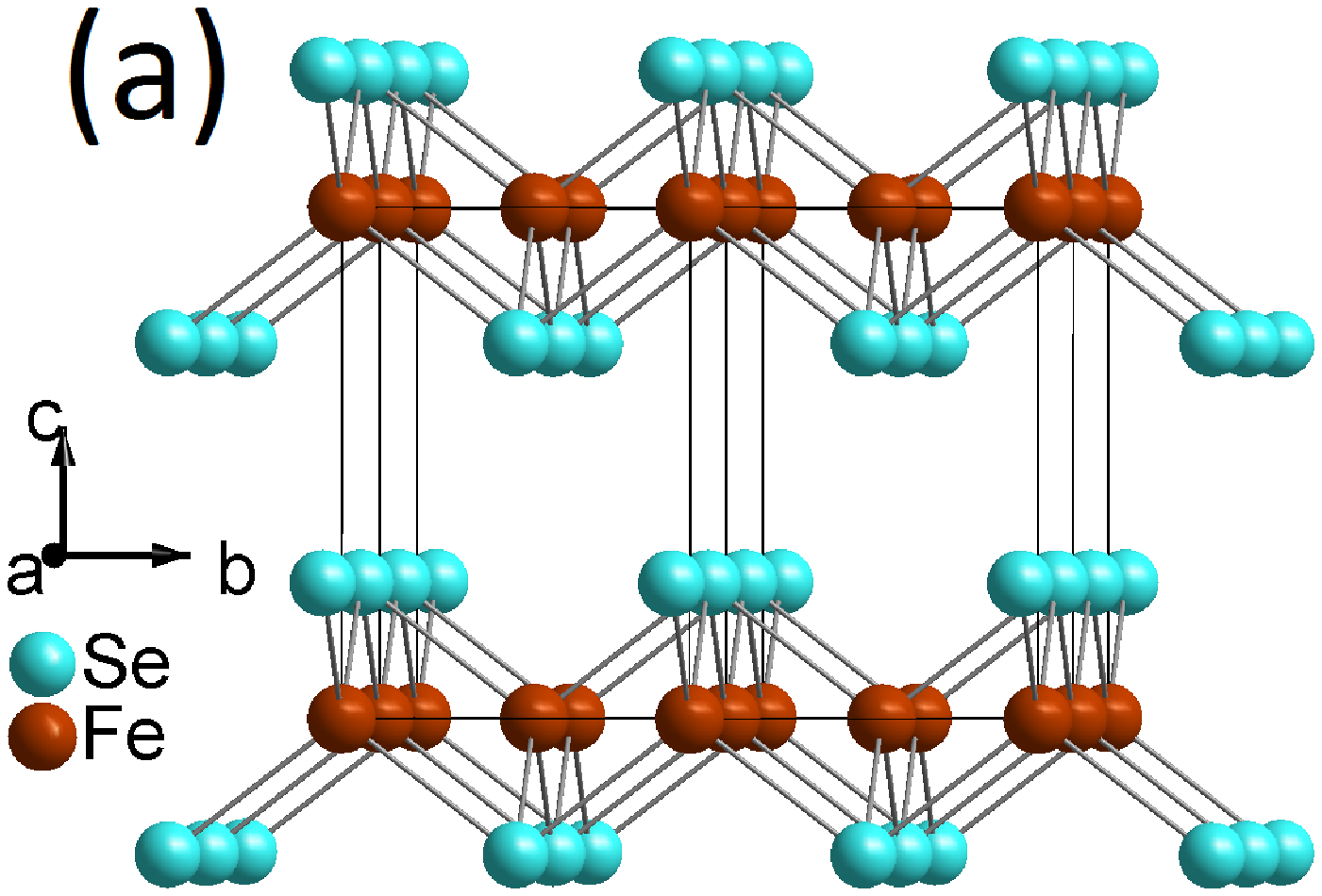}
\includegraphics[width=2.5 cm,clip]{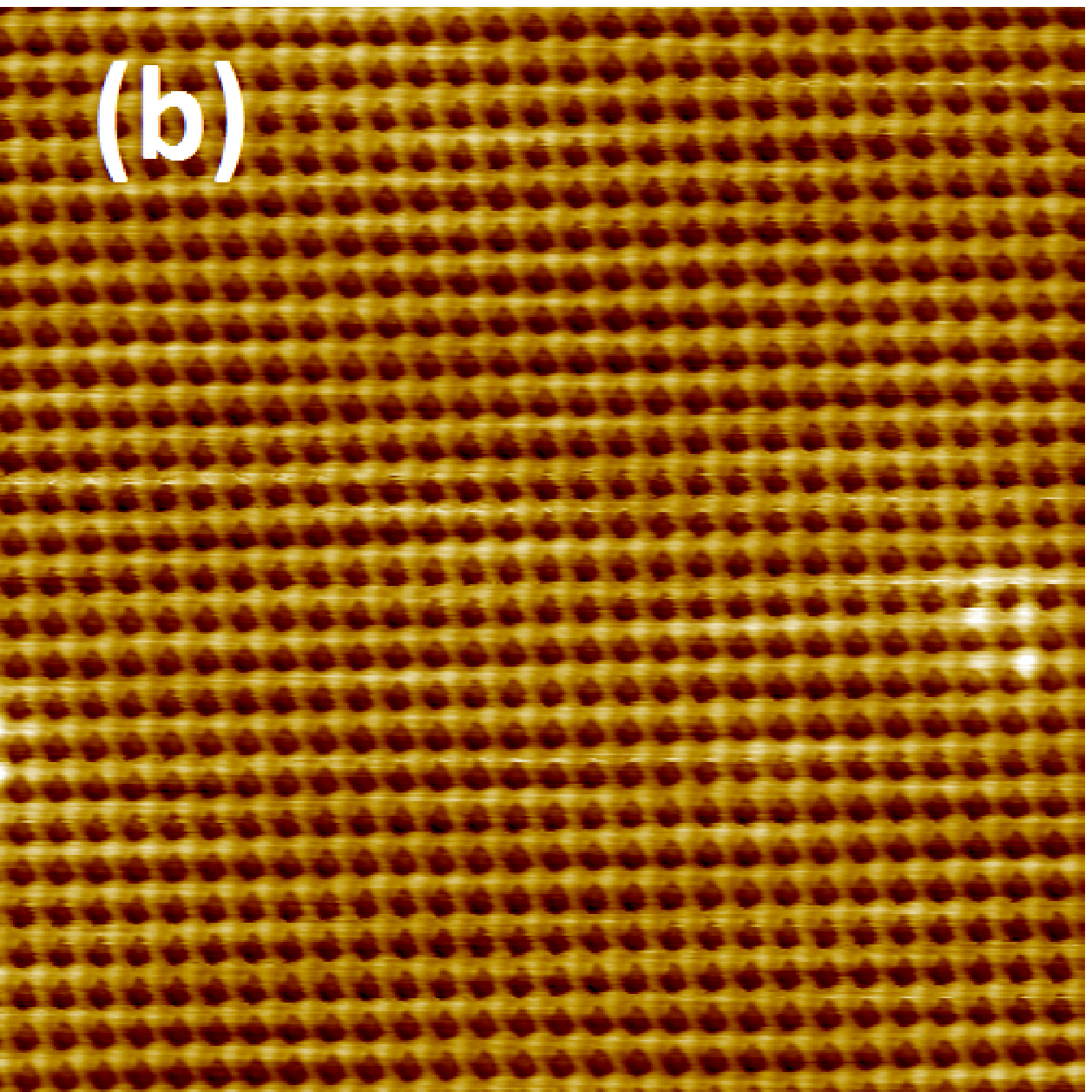}
\end{minipage}
\begin{minipage}{0.32\textwidth}
\includegraphics[width=5.5 cm,clip]{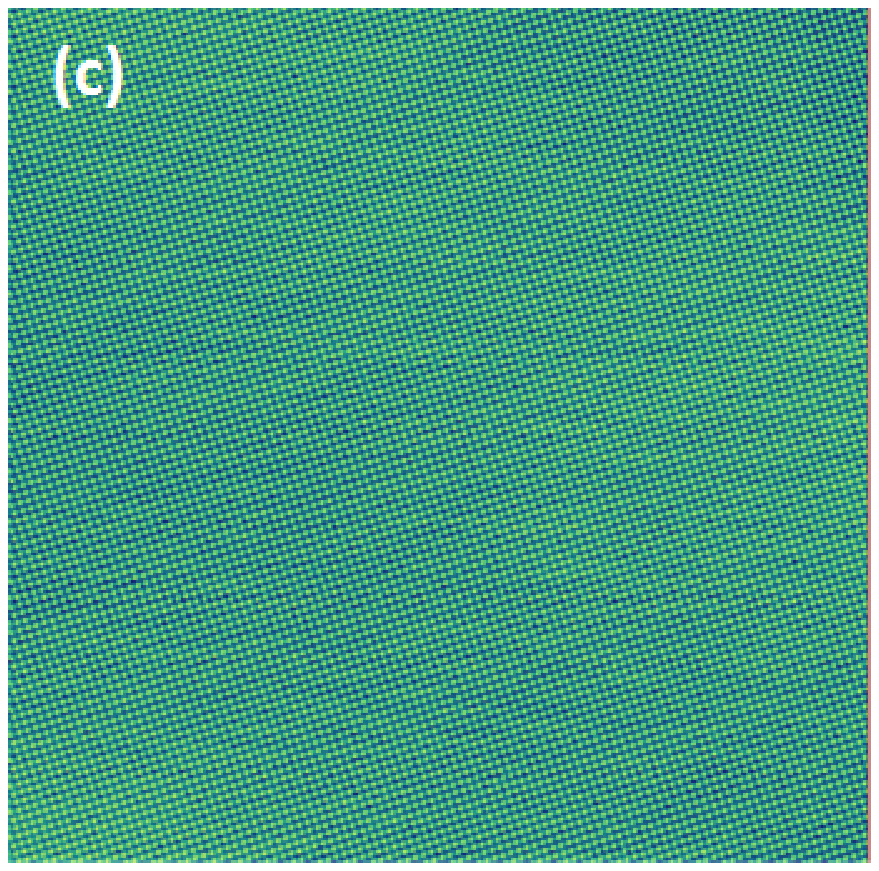}
\end{minipage}
\caption{(a)Tetragonal crystal structure of FeSe. (b) STM topography on an area of 10~nm $\times$ 10~nm, and (c) a topographic image over a larger area of 40 nm $\times$ 40 nm,  completely free of impurities and defects. The bias voltage and the tunneling currents in both cases were set at $V_{g}$ = 0.02~V  and $I_{sp}$ = 0.6~nA.}
\label{Fig1a}

\end{figure}

Our experiments were conducted on single crystalline FeSe grown by chemical vapor transport by taking a mixture of powdered FeSe and transport agent AlCl$_{3 }$ in the ratio 50:1. By this method, a layer-by-layer growth along the $c$-axis was achieved. 
 Our samples studied here displayed expected behavior of bulk superconductivity with  $T_{c}\approx$~8.5~K, and a structural transition at $T_{s}\approx$~87~K. A complete characterization proving a high quality of the crystals used here can be found in Ref. \cite{Koz2014}. The samples can easily be cleaved along the (001) plane. A total of four samples were cleaved in an ultra-high vacuum (UHV) chamber at around 20~K for STM experiments. The surface termination after cleaving contains a charge-neutral Se layer and hence, it is expected to provide an ideal, non-reconstructed surface for STM investigations. The tunneling spectroscopy was performed in the temperature range 6 - 95~K. The tunneling spectra were typically measured over an area of 5 nm $\times$ 5 nm, with a pixel resolution of 1~nm. The resistivity, transverse magnetoresistance (MR), and Hall-effect measurements in magnetic fields up to 9~T were conducted in a physical property measurement system (Quantum Design) with current along the tetragonal $a$-axis and magnetic field applied along the $c$-axis. The magnetization measurements were performed in a SQUID magnetometer (Quantum Design) with field applied parallel to the  $ab$-plane.
\begin{figure}[t]
\includegraphics[width=8.5 cm,clip]{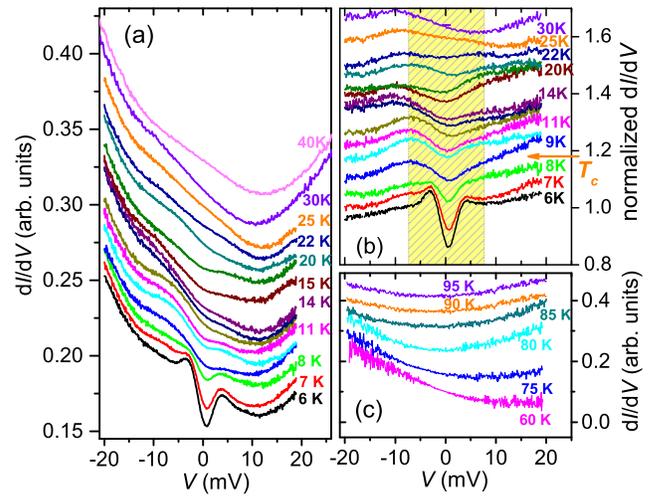}
\caption{(a) Raw tunneling spectra in the temperature range 6 - 40~K. (b) Normalized tunneling spectra in the temperature range 6 - 30~K. The yellow shaded region represents the energy scale of $T^{**}$. (c) Raw tunneling spectra in the temperature range 60 - 95~K. The spectra are shifted vertically for clarity. For all the spectroscopy measurements, $V_{g}$ = 0.02 V  and $I_{sp}$ = 0.6 nA before opening the STM feedback loop. The bias modulation amplitude was set to 0.1~m$V_{rms}$.}
\end{figure}
The topography of an \textit{in situ} cleaved FeSe single crystal revealed atomically flat and largely defect-free Se-terminated (001) surfaces, Fig. 1(b) and (c). The Se-Se distance \textit{a$_{Se\mathrm{-}Se}$}=  3.7($\pm$0.1)~\AA~observed at $T$ = 6~K is in good agreement with the distance of 3.7702(5)~\AA~ found in X-ray diffraction on our crystals. Note that, because the orthorhombic distortion in FeSe is less than 0.5\%, the expected rhombus structure of the Se-terminated surface can not be resolved from the STM topography.  

Spectroscopy conducted by means of an STM is a powerful technique if small energy gaps of the order meV at $E_\mathrm{F}$ are to be investigated. This is because the differential tunneling conductance $g(V)$= d$I(V)$/d$V$ is, within simplifying approximations, proportional to the DOS at $E_\mathrm{F}$. The area-averaged spectra $g(V)$ obtained in the temperature range 6 - 40~K are presented in Fig. 2(a). The $g(V)$ curves display a strong asymmetry for positive and negative bias voltages, which indicates an uncompensated nature of the electronic structure at these low temperatures. The higher values of $g(V)$ for the negative bias voltages indicate more electron-like carriers around $E_\mathrm{F}$. At 6~K, $i.e.$, only about 2.5 K below $T_{c}$, the superconducting gap as well as the accompanying coherence peaks are visible, particularly after normalization, Fig. 2(b). Since the $g(V)$ obtained at 40~K, $i. e.$, well above $T_{c}$, is essentially featureless in the low energy regime, it is  used for normalizing the $g(V)$ data collected at 6~K. A symmetrized and normalized 6~K spectrum was fitted to the Dynes gap function \cite{Dynes78,Hoff2011}. Because the Dynes function fails to describe all the spectral features (see Fig. S1 in \cite{Sup}) and since the position of the coherence peaks overestimates the gap magnitude due to the effects of thermal smearing on the tunneling spectroscopy at 6~K, we determined the gap size 2$\mathit{\Delta}_{6\mathrm{K}} = 3.51\pm0.02$ meV from the full-width at half-minimum of the symmetrized curve. This yields 2$\mathit{\Delta}_{6\mathrm{K}}$/k$_{\mathrm B}T_{c}$ = 4.76. We estimate $\Delta (T\rightarrow 0)\approx 2.19$ meV in close agreement with the value reported for FeSe single crystals \cite{Mal2014,Wata2015} as well as thin films \cite{Song2011,Song2014}.

\begin{figure}[t]
\centering 
\includegraphics[width=8.5 cm,clip]{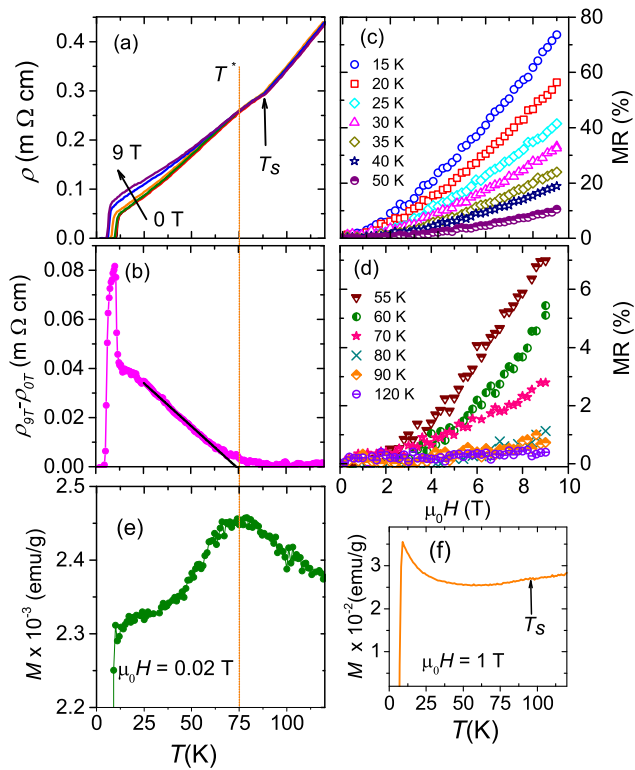}
\caption{(a) Resistivity $\rho(T)$ measured in magnetic fields of 0, 1, 3, 5, 7, and 9 T. (b) The difference between $\rho(T)$ measured in a field of 9~T and 0~T. The curve displays a change of slope at $T^{*}\approx 75$~K as shown by the black line. The magnetic field dependence of the magnetoresistance MR in the temperature range (c)15 - 50~K and (d) 55 - 120~K. (e) and (f) Magnetization $M(T)$ measured for different applied fields after zero-field cooling the sample.  $T^{*}$ represent the onset temperature of the spin fluctuations.}  
\label{Fig3}
\end{figure} 
Upon increasing temperature, two features can be discerned from the temperature evolution of   $g(V)$: ($i$) the superconducting gap and coherence peaks disappear at around 8-9~K, as expected from  $T_{c}\approx$ 8.5~K. ($ii$) a gap-like suppression in the DOS at $E_\mathrm{F}$  persists well above $T > T_{c}$; a corresponding hump at about -8~meV can easily be traced at least up to 20 K in Fig.~2(a, b). Since the bottom (top) of the electron (hole) band is found to be only a few meV away from $E_\mathrm{F}$ \cite{Mal2014,Kas2014}, the width of this suppression is difficult to estimate. Further, the $g(V)$ as measured here is not sensitive to the momentum. Hence, at present it is not clear which part of the electronic band structure is suppressed, or may be even gapped, in momentum space. Above $T\approx 22$~K, features in the spectra fade away due to the effects of thermal broadening, as inferred from Figs. S2 \cite{Sup}. This implies that 22~K is the lower limit for the onset temperature of the partial gap that can be determined from STM spectroscopy data.


%
To address the origin of the electron-hole asymmetry in the band structure, the temperature dependence of $g(V)$ was investigated up to $T>T_{s}\approx 87$~K. In Fig. 2(c), the tunneling spectra in the temperature range 60 - 95~K are presented. Above $T^{*}\approx 75$~K, the tunneling spectra tend to become symmetric indicating that the electron-hole asymmetry observed at lower temperatures is likely related to enhanced spin fluctuations. Such enhanced spin fluctuations are known to exist below $T_{s}$ and they will be discussed below. 
 
    In order to examine whether the above mentioned conjecture is correct, we performed electrical transport and magnetization measurements. This also ensures that the properties we observe by STM are related to the bulk properties of the samples. The temperature dependence of the resistivities $\rho(T)$ in different magnetic fields are presented in Fig. 3(a). $\rho(T)$  displays an overall metallic behavior with a small anomaly at the structural transition marked by $T_{s}$. The magnetoresistance MR =  $[\rho(H)-\rho(0)]/\rho(0)$ is positive below $T_{s}$. In Fig. 3(b), the difference between $\rho(T)$ measured in a field of 9~T and 0~T is plotted. This curve displays a change of slope at a temperature $T^{*}\approx$ 75~K. Note that, in our interpretation, $T^{*}$ is not a well defined temperature, but rather a cross-over energy scale. This is supported by Fig. 3(b).  Below $T^{*}$, the MR is dramatically enhanced as seen also in Figs. 3(c) and (d), where the field dependencies of the MR are plotted. For $T>T^{*}$, however, the MR is about 1\% (Fig. 3(d)), which is an expected behavior of conventional metals.  Figs. 3(e) and (f) present the magnetization $M(T)$ measured in fields of 0.02~T and 1 T, respectively. A distinct down-turn observed at 75~K in Fig. 3 (e) suggests an onset of antiferromagnetic spin-fluctuations. The magnetization measured in a field of 1~T presented in Fig. 3(f) shows a tiny anomaly at  $T_{s}\approx87$~K(marked by an arrow in Fig. 3(f)). In contrast to the low-field case, however, the magnetization displays a broad minimum and then increases upon further decreasing the temperature, before entering into the superconducting phase at  8.5~K. 
As expected, $M(T)$  measured at 1~T resembles the temperature dependence of the spin-lattice relaxation rate $1/T_{1}T$  measured in the NMR studies of FeSe  \cite{Baek2014,Bom2015,Imai2009}. The strong enhancement of the positive MR below $T^{*}$ can thus be related to a $T$-dependent anisotropic scattering of the charge carriers induced by the spin fluctuations below $T^{*}$. 	

The above picture is also confirmed by the scaling behavior of the MR. If the scattering rates for charge carriers are isotropic, $i.e.,$ equal at all points of the Fermi surface, the MR  should scale with magnetic field $H$ via an arbitrary function $\mathcal{F}[H/\rho(0)]$  regardless of the topology of the Fermi surface, a behavior known as Kohler's rule \cite{Pip1989}.  In Figs. 4(a) and (b), the Kohler plot for FeSe in the temperature range 20 - 120~K is presented.  All the MR data at $T \leq $ 30~K scale onto a single curve. This allows us to define a second energy scale $T^{**} \approx 22 -30$~K, a temperature below which Kohler's rule is valid and a partial gap is observed in STS. However, the scaling is not valid, $i.e.,$ Kohler's rule is violated, for the range  30~K $< T \leq$ 70~K suggesting anisotropic scatterings of the charge carriers in this temperature range. 	For temperatures above $T^{*}$, as seen in Fig. 4(b), the scaling of MR is obeyed again, in line with the statement above that FeSe behaves as a conventional metal at these temperatures. Thus, our results suggest that the anisotropies of the Fermi surfaces are associated with the spin fluctuations rather than the structural phase transition at a higher temperature $T_{s}\approx87$~K. 

The validity of Kohler's rule below $T^{**}$ can be explained by considering nesting driven spin or charge density wave (CDW) fluctuations. The quasiparticle scattering by these density wave fluctuations is momentum dependent \cite{Kemp2011,Brei2014} as indicated by a peak of the spin/charge susceptibility in momentum space along the possible ordering (nesting) vectors {\bf Q}. Those parts of the Fermi surfaces  that are connected by the ordering vectors {\bf Q} are known as hot spots, where scattering is expected to be strong, and  quasiparticle life-times are short. The converse is true for the cold parts of the Fermi surface. 	Violation of  Kohler's rule supports the existence of hot spots and cold parts in momentum space with anisotropies in their scattering rates, yet both contributing to the transport. However, below $T^{**}$,  the MR data collapse onto a single curve indicating the suppression of anisotropic scatterings due to opening up of a gap,  presumably at the hot spots of the Fermi surface. Thus, the results of the transport measurements are highly consistent with those of the tunneling experiments. 
 
\begin{figure}
\includegraphics[width=8.5 cm,clip]{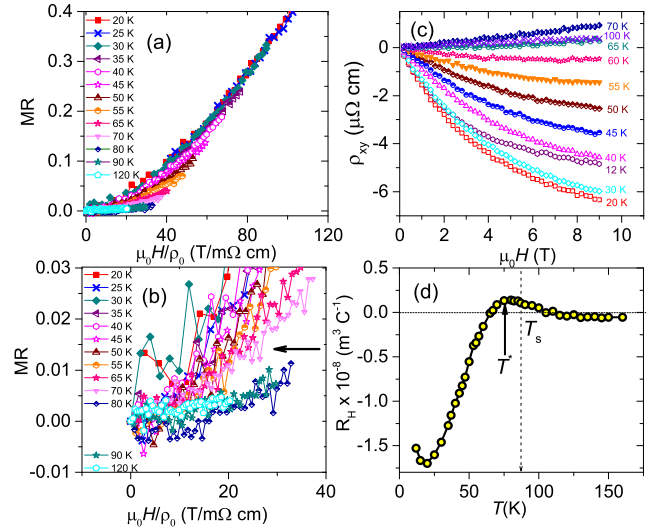}
\label{fig4}
\caption{(a) Kohler plot in the temperature range 20-120 K displaying the validity of Kohler's scaling below 30~K and above 70~K. (b) Same data shown is panel (a), but on an enlarged scale. The horizontal arrow is placed to emphasize that $T^{*}$ is different than $T_{s}$.  (c) Hall resistivity $\rho_{xy}$ in the temperature range 12-100~K. (d) The temperature dependence of the initial Hall coefficient $R_{H\rightarrow0}$ displays a clear deviation from the compensated metal regime below $T^{*}\approx$ 75~K where spin fluctuations become important.} 
\end{figure}

The Hall effect measurements on FeSe are also in accordance with the above interpretation. In  Fig. 4(c),  the Hall resistivities, $\rho_{xy}(H)$, measured at different temperatures are presented.
The initial Hall coefficient $R_{H\rightarrow0  }(T)=\frac{\delta \rho_{xy}(T)}{\delta H}\rvert_{H\rightarrow0}$ is given in Fig. 4(d).  $R_{H\rightarrow0  }(T)$ does not show a significant change at $T_{s}$, but goes through a maximum at $T_{max}^{Hall} \approx 75$~K = $T^{*}$, and subsequently becomes negative below 60~K. Note that it can be seen from Fig. 4(d) that, at $T^{*}$,  the system initiates to undergo a transition from a nearly compensated metal to a strongly electron dominated transport regime. These results are consistent with the strong electron-hole asymmetry found in the tunneling spectra below 75~K.

The cross-over phenomenon observed at $T^{*}$ signals the appearance of an incipient, but likely incomplete ordering mode. This is associated with an anisotropic evolution of the Fermi surface, as seen from the violation of Kohler's rule. Below $T^{**}$, both, the validity of Kohler's rule as well as the direct visualization of a partial gap in the local DOS suggest that the fluctuations of the ordering mode become more static upon reducing temperature. The opening up of the partial gap in the DOS points a scenario of a subtle onset of some type of charge or orbital density wave that breaks the translational invariance of the lattice. Hence, it is clearly distinct from the nematic phase transition at which only the rotational symmetry of the lattice is broken. In  the nematic state, the occupancies $n$ of the $d_{xz}$ / $d_{yz}$ orbitals are not equal \textit{i.e.}, $n_{xz}\neq  n_{yz}$. This may trigger a Peierls-type dimerization. An early electron diffraction experiment \cite{Mc2009} on FeSe indeed reports that the actual symmetry of FeSe below 20~K is lower than the $Cmma$ space group. Since the dimerization is expected to occur in the Fe-layer, its effect can not be directly discernible in the Se-terminated STM topographic image. The telltale sign of the ordering mode in both spin as well as charge channels observed here indicate that FeSe is a spin-charge coupled system. Theories for Fe-SCs based on Fermi surface nesting predict possibilities of various types of spin-, charge-, and orbital/pocket-density waves (SDW, CDW, PDW, respectively), as well as their mutually orthogonal linear combinations \cite{Jian2011}.    
				
In conclusion, our STM, magnetization, and transport data consistently show evidence for an onset and a nucleation of an ordering mode. Although no long-range magnetic order is observed in FeSe, our results strongly suggest that spin fluctuations still play an important role in determining the transport properties. The results suggest an existence of some type of density wave, while its exact nature remains an open problem. However, the spin degrees of freedom are coupled with it.

\paragraph* {Acknowledgments:} We are indebted for stimulating discussions with S.-H. Baek, S. Borisenko, A. V. Chubukov, A. Coldea, D. V. Efremov, J. Fink, Y. Grin, M. Haverkort, Z. Hu, D. Kasinathan, D. I. Khomskii, M. Nicklas, H. Rosner, Q. Si, U. Stockert,  P. Thalmeier, L.~H.~Tjeng, and A. A. Tsirlin. Financial support from the Deutsche Forschungsgemeinschaft (DFG) within the Schwerpunktprogramm SPP1458 is gratefully acknowledged.

\end{document}